\documentclass[acmlarge]{acmart}

\AtBeginDocument{%
  \providecommand\BibTeX{{%
    \normalfont B\kern-0.5em{\scshape i\kern-0.25em b}\kern-0.8em\TeX}}}

\setcopyright{acmcopyright}
\acmJournal{IMWUT}
\acmYear{2020} \acmVolume{4} \acmNumber{3} \acmArticle{114} \acmMonth{9} \acmPrice{15.00}\acmDOI{10.1145/3414173}

\acmConference[UbiComp/ISWC '20]{UbiComp/ISWC '20}{September 12--16, 2020}{Cancún, México}

\usepackage[german, polish, english]{babel}
\usepackage[utf8]{inputenc}
\addto{\captionsenglish}{\renewcommand{\bibname}{References}}

\usepackage{todonotes}
\usepackage{caption}
\usepackage{tabularx}

\usepackage{soul}
\soulregister\cite7
\soulregister\ref7
\soulregister\pageref7

\usepackage{xcolor}

\usepackage{array, multirow}
\newcolumntype{C}[1]{>{\centering\arraybackslash}m{#1}}

\begin{document}
\sethlcolor{green}
\renewcommand{\bibname}{References}

\title[SAFER: An IoT Device Risk Assessment Framework]{SAFER: Development and Evaluation of an IoT Device Risk Assessment Framework in a Multinational Organization}

\author{Pascal Oser}
\email{p.oser@cern.ch}
\orcid{0000-0002-2461-343X}
\affiliation{%
  \institution{Ulm University}
  \city{Ulm, Germany}
}
\affiliation{     
\institution{European Organization for Nuclear Research (CERN)}
  \city{Geneva, Switzerland}
}

\author{Sebastian Feger}
\email{sebastian.stefan.feger@cern.ch}
\orcid{0000-0002-0287-0945}
\affiliation{%
  \institution{Ludwig Maximilian University of Munich}
  \city{Munich, Germany}
}
\affiliation{     
\institution{European Organization for Nuclear Research (CERN)}
  \city{Geneva, Switzerland}
}

\author{Paweł W. Woźniak}
\email{p.w.wozniak@uu.nl}
\affiliation{%
  \institution{Utrecht University}
  \city{Utrecht, Netherlands}
}

\author{Jakob Karolus}
\email{jakob.karolus@ifi.lmu.de}
\affiliation{%
  \institution{Ludwig Maximilian University of Munich}
  \city{Munich, Germany}
}

\author{Dayana Spagnuelo}
\orcid{0000-0001-6882-6480}
\affiliation{%
  \institution{Vrije Universiteit Amsterdam}
  \city{Amsterdam, Netherlands}
}

\author{Akash Gupta}
\email{aksgupta3697@gmail.com}
\orcid{0000-0002-1499-6123}
\affiliation{%
  \institution{National University of Singapore}
  \city{Singapore}
}

\author{Stefan Lüders}
\email{stefan.lueders@cern.ch}
\orcid{0000-0001-8676-2353}
\affiliation{     
\institution{European Organization for Nuclear Research (CERN)}
  \city{Geneva, Switzerland}
}

\author{Albrecht Schmidt}
\email{albrecht.schmidt@ifi.lmu.de}
\orcid{0000-0003-3890-1990}
\affiliation{%
  \institution{Ludwig Maximilian University of Munich}
  \city{Munich, Germany}
}

\author{Frank Kargl}
\email{frank.kargl@uni-ulm.de}
\orcid{0000-0003-3800-8369}
\affiliation{%
  \institution{Ulm University}
  \city{Ulm, Germany}
}

\renewcommand{\shortauthors}{Oser, et al.}

\begin{abstract}
Users of Internet of Things (IoT) devices are often unaware of their security risks and cannot sufficiently factor security considerations into their device selection. This puts networks, infrastructure and users at risk. We developed and evaluated SAFER, an IoT device risk assessment framework designed to improve users' ability to assess the security of connected devices. We deployed SAFER in a large multinational organization that permits use of private devices. To evaluate the framework, we conducted a mixed-method study with 20 employees. Our findings suggest that SAFER increases users' awareness of security issues. It provides valuable advice and impacts device selection. Based on our findings, we discuss implications for the design of device risk assessment tools, with particular regard to the relationship between risk communication and user perceptions of device complexity.
\end{abstract}



\maketitle

\section{Introduction}

The number of connected devices deployed in home and office environments increased rapidly over the past years and is expected to further increase in the future \cite{Gartner2019Growth_Of_Connected_Devices}. This process is driven by the development of Internet of Things (IoT) devices that often do not adhere to technology recommendations (e.g., ENISA \cite{ENISA2017IoT_recommendations, ENISA2019Standards_comparison}) or structured support schemes---at the expense of the security of those accessories. Vulnerabilities have been detected in devices ranging from CCTV systems \cite{Darkreading2018vulnerable_CCTVs} to most intimate gadgets \cite{Wired2019IoT_sex_toys, LATimes2016IoT_privacy}. Despite such alarming reports, users often do not---or cannot---sufficiently consider security criteria in the selection and during the use of networked devices. While the growing number of connected devices enable powerful new use cases, they introduce vulnerabilities in networks and open them up for attacks that risk, at the very least, to jeopardize users' privacy. This is becoming a growing concern of the UbiComp community that recently investigated IoT device security schemes operating without user intervention \cite{fomichev2019perils}; needs in control and accountability of IoT devices in the smart home \cite{jakobi2018evolving}; and privacy and control over intimate data recorded by connected devices \cite{kwon2018connected}.

In past years, computer security research has made progress in systematically evaluating risks and security of IoT systems \cite{Loi2017}. Yet, it becomes evident that the risk assessment of those devices is not just a technological issue, but also a social one. In fact, security ceremonies formally recognize and emphasize the human role in computer security \cite{bella2012layered}. Past research shows that users need to be involved in the evaluation of different risk assessments and representations \cite{Huang2016}.

Related work in the HCI community emphasized the need for research on the security of IoT devices that often have minimal security by design \cite{Poslad:2013:ASP:2494091.2499770}. The suitable presentation of device information is crucial to enable users to make informed decisions on the reliability and trustworthiness of IoT accessories \cite{Pignotti:2013:TTT:2523501.2523503}. In the spirit of recent calls for HCI scholars to impact IoT systems development with "design for responsibility" \cite{Fritsch:2018:CRA:3173574.3173876}, our study focuses on understanding the impact of IoT device risk communication on users' security awareness and assessment. To do so, we developed and deployed SAFER, an IoT device risk assessment framework. We deployed SAFER in a large multinational research organization, namely the European Organization for Nuclear Research (CERN), that employs around three thousand professionals\footnote{\url{https://cds.cern.ch/record/2719035/files/CERN-HR-STAFF-STAT-2019-RESTR.pdf}}, accommodates ten thousand visiting researchers\footnotemark[1], and manages references to more than 300,000 registered devices --- many under a "Bring Your Own Device" (BYOD) policy. This focus on a professional organizational offers a new perspective on employees' IoT security considerations. So far, related work focused mostly on security factors in purely private environments. Here, security is not yet a primary consideration in device selection, but easily understandable and trustworthy security assessments are expected to impact consumers' security awareness \cite{emami2019exploring}. While device use in private home networks can compromise users' intimate environments, vulnerabilities in organizational environments can risk critical infrastructure and lead to administrative sanctions. 

We implemented two device risk assessment views in SAFER. One version offers a \textit{guided} overview of device risks and proposes actions to improve the security of the devices or to protect the network from security flaws. The other offers a \textit{rich} version, which provides detailed information about the evaluated data. We conducted a between-subjects, mixed-method study with 20 employees to understand how those different representations impact users' security awareness and acceptance of device sanctions. 
In order to profit from the shared business-private network environment, we asked the study participants to review both private and business devices. Participants further studied information on SAFER as part of a selection process for one new smartphone, representing the private device, and one new network-attached storage device, representing the business device.

Our paper makes the following contributions:

\begin{itemize}
    \item 
    We map practices around IoT device usage and security factors in a multi-national organization with a BYOD policy across a diverse set of employees with different professional backgrounds and roles;
    
    \item 
    We report on the development of SAFER and our user study that contrasted different device risk assessments (i.e. Low / Medium / High), two device risk representation views (i.e. Guided vs. Detailed), and two device categories (i.e. business / private devices);
    \begin{itemize}
    \item 
    We contribute a comparison of users' security perceptions in the form of willingness to monitor devices, concerns for devices, and device removal acceptance in relation to the device risk assessment and the device categories;
    
    \item 
    We compare the perceived functionality, helpfulness, and device assessments of the two versions and provide qualitative insights explaining how SAFER supports risk assessment;
    \end{itemize}
    \item We present implications for the design of technology that enables users to take informed decisions about device risks, sanctions, and selections in the IoT context.
\end{itemize}

This paper is structured as follows. First, we reflect on risk communication and the role of security in device selection. Second, we depict socio-technical aspects of IoT and computer security. Third, we detail the implementation of the SAFER framework and the two device risk assessment views. Fourth, we provide an extensive description of the study participants and the study protocol. Next, we present the results and findings of the evaluation. Finally, we discuss design implications for IoT device risk assessment tools that are essential for ubiquitous environments and the UbiComp community.

\section{Related Work}

Poslad et al. \cite{Poslad:2013:ASP:2494091.2499770} remarked the shifting meaning and definition of the Internet of Things (IoT). Initially, it was conceptualized as a network of specific things, designed to enrich information about the world and to interact with its environments. As the authors point out, "this vision has since expanded to include a more diverse range of devices, services and networks to become an Internet of anything, anywhere, connected, anyhow." Following this notion of the Internet of \textit{anything}, we stress the broader context in which we consider IoT in this study: as any \textit{connected device} that interacts with infrastructure of the organization. In this section, we reflect on device risk assessments and risk presentation. We reflect on usable security research that focuses on users' device security considerations in private environments. We relate those study designs and findings to our research in a professional BYOD network.

\subsection{Risk Assessment and Presentation}

Computer security research progressed in systematically evaluating the security of IoT devices. Recent work demonstrated that IoT devices can be categorized and identified in a network through machine learning algorithms \cite{oser2018identifying}. This is a first crucial step on the way to large-scale, automated risk evaluations. Another major challenge is the design of suitable and accessible risk presentation methods, as Huang et al. stressed \cite{Huang2016}. They developed a security framework for IoT devices that offers three basic risk representation methods: risk table, risk tree map and risk tree, where device threats are the tree's children and leafs represent risks. Based on an evaluation with 12 participants, they found significant differences between those risk representations in terms of easiness of understanding and preference.  

In fact, that approach relates to Emami-Naeini et al.'s work \cite{emami2019exploring}. In order to make device privacy and security information more accessible, they designed paper prototype labels for three smart devices and three label variants for each of those devices. The labels contained privacy and security information and were found useful and accessible by the study participants. However, the participants stressed that trust in the security information and the provenance of those information is highly important. 

In their discussion of trust in a specific technology, McKnight et al. \cite{mcknight2011trust} distinguish between knowledge-based trust in technology and initial trust. They stress that "when individuals rely on knowledge-based trust, they draw less on institution-based beliefs, and make decisions based on trusting beliefs about characteristics of the technology itself." Related to the previously discussed context of trust in the provenance of security information, McKnight et al. note that it is necessary to examine "the dynamic interplay between users' trust in human agents that built a system, human agents that introduce a system, those that support a system, and the technology itself." In order to investigate trust in security evaluation technology and the developers of such technology, \textit{we employed several of their proposed scales in an IoT context}.

Chuang et al. \cite{chuang2018design} proposed a design vocabulary for human-IoT device communication that expects users to better anticipate the status of IoT devices. Loi et al. \cite{Loi2017} also emphasized the value of suitable and correct risk communication to end users. They categorized IoT device risks along four dimensions: "\textit{confidentiality} of private data sent to/from the IoT device; \textit{integrity} of data from the IoT device to internal/external entities; \textit{access control} of the IoT device; and \textit{reflective attacks} that can be launched from an IoT device." The authors developed software designed to automate the security testing within these four domains. They present the security evaluation of a device as a color-coded rating table based on 24 rated items that cover the four dimensions. The authors envision the development of a rating system that is more accessible to end consumers. Our study relates to this vision, as SAFER is design to offer visual ratings and detailed information related to the development of risks and its contributing factors.

Caivano et al. \cite{Caivano:2018:TIM:3206505.3206587} remarked that IoT devices can be used in different environments, from home to industry. They further stressed that end users were no technical experts. Thus, tools need to be developed that enable them to control their devices better. Their initial proposal of an IoT model for device assessments aims at "facilitating the choice of the devices that better suit the domain in which they should be used." While this model captures four dimensions --- communication, target, data manipulation and development --- security considerations are limited to communication cryptography. Our work aims to contribute to any such developments of an IoT model that entails a rich set of security factors which help users make informed decision in the selection of their devices.

\subsection{The Role of Security in Device Selection}

Emami-Naeini et al. \cite{emami2019exploring} studied how privacy and security affect consumers' IoT device purchase behaviors. They interviewed 24 participants who had purchased IoT accessories. As part of their semi-structured interview study, they asked participants about the types of devices that they purchased and the factors that influenced purchase decisions. They further asked about any prior decisions not to buy a device, and about post-purchase concerns and the way those concerns were managed. Most of the participants reported that privacy and security were not considered as part of the device selection process. However, those consumers got worried through media reports later. Instead, the participants "who sought privacy and security information before purchase, reported that it was difficult or impossible to find."

Our study aims to understand, how security factors into the use and selection of both private and business devices in an organizational environment, and how technology impacts security awareness and evaluation. To relate our findings to the study on purely private devices and private networks by Emami-Naeini et al., \textit{we re-used and adapted several parts of their openly available interview protocol}. Especially those parts about the selection and concerns related to connected devices. Instead of paper prototype labels, we used SAFER, a device risk assessment framework that we developed. Doing so, we aimed to relate our findings from a mixed private-business work environment to findings from private smart device environments.

\section{Socio-Technical aspects of security}

IoT devices are diverse and they are not always designed with security as a priority. Devices ranging from Programmable Logic Controllers (PLCs), smart thermometers and smart appliances, CCTV cameras, and printers are only a few examples of general purpose IoT devices that can be found in a large organization \cite{oser2018identifying}. These devices are highly heterogeneous with regard to the communication protocols used, and the range of other devices they can control or be controlled by  \cite{ding2019ethical}. If IoT devices can obtain sensitive data from businesses and individuals, attacks on IoT vulnerabilities can put the entire network at a security risk \cite{girma2018analysis}. 

Intuitively, one could imagine that the solution for that is purely technical; it lies on inspecting these devices individually, implementing protective measures in the network, and applying security sanctions. However, in organizations with Bring Your Own Device (BYOD) policies, it would be unwise to overlook the role humans take on the overall security of the organization (see, for instance, the security and privacy considerations made by Miller et. al. \cite{miller2012byod}). Employees, unaware of the possible vulnerabilities on their IoT devices, will keep introducing new security risks to the network.
Following this reasoning, in a multinational organization, thorough security can only be achieved if approached from a \textit{social} and \textit{technical} perspective.

The role of humans in security is already recognized in computer security ceremonies \cite{bella2012layered}, which come to extend the purely technical approach to security. Security ceremonies incorporate human interaction and the environment influences in its formal analysis. It also incorporates the idea that different devices are designed with their own environment and security assumptions in mind, and when put to work together they can lead to unpredictable security flaws \cite{martina2010should}. 

Kirlappos and Sasse \cite{kirlappos2014usable} reflect on the role employees play in organizational security. They "argue that an important but often-neglected aspect of compliance is trusting employees to 'do what's right' for security." The authors stress that trust needs to be incorporated into security design. They discuss four key elements that require improvements, of which two are particularly relevant in the context of our research: \textit{participation}, here with the meaning of the "organizational ability to identify problems"; and \textit{usability}, which in the context of security enables the employee to behave in a trustworthy manner. The authors also point out the role of security \textit{awareness}, in which employees should be reminded of the critical role they play in protecting the organization and its resources.

Concepts of awareness arguably have also been subject of research in IoT devices.
Fritsch et al. \cite{Fritsch:2018:CRA:3173574.3173876}, for instance, analyze IoT manifestos which describe \textit{responsibility} as key challenge and concern. In here, responsibilization is discussed as a term related to \textit{understanding} and having the ability to assess the design, data management and potential consequences of IoT technology use. They stress that "responsibility for conveying information about the device is turned into a design problem, one of condensing information that would make sense to consumers." The authors call for HCI to "theorize and design for responsibility while attending to the perils of responsibilisation."

Responsibilization refers to the transfer of responsibility to the actual device users and owners. Pignotti and Edwards \cite{Pignotti:2013:TTT:2523501.2523503} stress that in order for consumers to take any responsibility during device selection, use and maintenance, the IoT needs to become more transparent. They investigated how Semantic Web technologies can help to manage information about IoT devices, and how an effective management of that information "would allow users to make informed decisions on the trustworthiness of such devices based on their provenance and use."

Our study aims to contribute to design for responsibility. The SAFER framework is designed to bring awareness to employees about security risks by providing comprehensible explanations of vulnerabilities and recommended actions.

\section{The SAFER System}

SAFER is an IoT device risk assessment framework we developed to improve users' ability to assess the security of connected devices.
In this section, we describe the architecture of the framework and its deployment in a large multinational organization. 
We further detail the design of two device risk assessment presentations.
\begin{figure*}
  \centering
  \includegraphics[width=1.0\columnwidth]{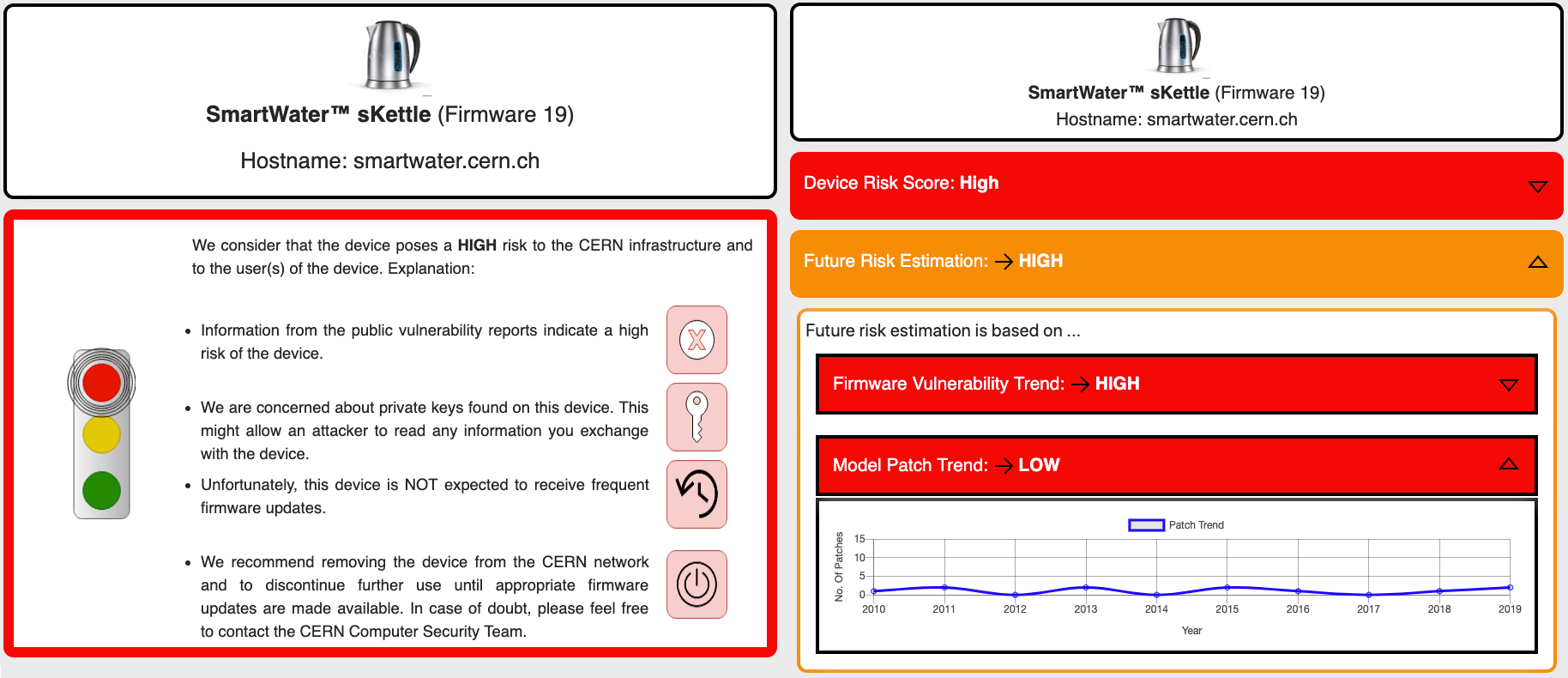}
  \caption{SAFER provides two device risk assessment views. The guided view (left) focuses on a traffic light-based security assessment and explanatory text. The detailed view (right) surfaces all data related to the security assessment. It is designed to aid users through tool-tips that explain every single data element on this view. }~\label{fig:figure_dashboard_combined}
\end{figure*}

\subsection{General Architecture} \label{SAFER_Framework:General_Architecture}

SAFER is based on four core components: \textit{Device identification, vulnerability enrichment, scoring, and front-end}. 
To prevent attackers from tricking SAFER by taking over and mimicking a secure device, the \textit{device identification} component launches two different scan mechanisms to validate the identified device category, manufacturer, model, and possibly firmware version of queried devices. 
One identification mechanism we use from Oser et al. \cite{oser2018identifying} is based on TCP timestamps \cite{TCP_Timestamps}, which relies on the hardware of embedded devices. 
The second mechanism uses characteristic web patterns of a device's web-page to identify the device in question. 
Those include distinctive manufacturer or device model strings, copyright statements, images, application programming interfaces (API's), and firmware versions.

The identification component can easily be enhanced with more mechanisms due to a novel data fusion approach using Subjective Logic \cite{josang-subjective-2016} which SAFER utilizes to find the correct device model and firmware out of multiple identification sources.

The \textit{vulnerability enrichment} component extracts the firmware to find third-party software. It queries multiple public sources like Mitre \cite{Mitre2019CVE} for information on vulnerabilities associated with the third-party software and the specific device model. All queried vulnerabilities, along with exceptional risks of the firmware (e.g. if SAFER found private keys) are transmitted to the scoring component in the next step.

The \textit{scoring} component assesses the risk of devices based on the vulnerabilities and exceptional risks. The scoring component generates a \textit{firmware vulnerability trend} across all firmware images, which indicates how many vulnerabilities were introduced and patched over the past years. 
This trend relates to the criticality of vulnerabilities introduced by third-party software or by the manufacturer himself. As a result, the trend shows users the most common severity of all currently unpatched security vulnerabilities. SAFER defines this trend as low, medium or high based on the CVE severity score.
The \textit{model patch trend} provides an average of how long it takes a manufacturer to patch vulnerabilities in their firmware images once they become publicly known. SAFER calculates this trend by generating individual patch time-spans for every firmware image. For every detected vulnerability, SAFER measures the time from it being publicly registered in a vulnerability database until the manufacturer patching the vulnerable software. Depending on the vendor's activity, SAFER rates the trend as slow, medium or fast. This provides users with a robust estimate of how long their device on average is left vulnerable to security vulnerabilities by the vendor.
Combining the firmware vulnerability and model patch trends from past record, and extrapolating it to the future, results in the future risk estimation of this particular model. 

\begin{table}
\caption{Future Security Risk Levels}
\label{Table:futureSecurityRiskLevel}
\resizebox{\columnwidth}{!}{
\begin{tabular}{ |C{3cm}||C{3cm}|C{3cm}|C{3cm}|C{3cm}| }
 \hline & 
\multicolumn{1}{c}{} & 
\multicolumn{1}{|c}{} & 
\multicolumn{1}{c}{\large \textbf{Model Patch Trend}} &
\multicolumn{1}{c|}{} \\

 \hline \hline
  &  & \large \textbf{Fast} & \large \textbf{Medium} & \large \textbf{Slow}\\
 \hline
  \large \textbf{Firmware} & \large \textbf{Low} & \large Low & \large Low & \large Medium\\
  \large \textbf{Vulnerability} & \large \textbf{Medium} & \large Low & \large Medium & \large High\\
 \large \textbf{Trend} & \large \textbf{High} & \large Medium & \large High & \large Critical\\
 \hline
\end{tabular}}
\end{table}

SAFER estimates the future risk level based on a risk matrix defined in Table~\ref{Table:futureSecurityRiskLevel}. 
As an example, we define fast patching manufacturers leaving high rated vulnerabilities unpatched as \textit{medium future risk}. We argue that only a skilled attacker can exploit this vulnerability at this moment and it is likely to be patched in the near future due to the fast patch trend. This risk estimation assists and warns users of future risks that devices can expose to a network, even if latest firmware versions are applied immediately.

In the last step, the \textit{front-end} component displays all risk information of the current firmware, the different trends, the future risk estimation, and exceptional risks to the user. The visualization of these results are either displayed in a \textit{guided} or information-\textit{rich} view.

We evaluated the correctness of all above mentioned components by manually investigating and validating the results of SAFER. To do so, we took multiple representative devices of every device model and verified SAFER's components. We manually accessed the web-pages and compared the device identifications of both identification mechanisms. We evaluated the device identification approach by Oser et al.~\cite{oser2018identifying} on even more devices. After that, we reverse-engineered the firmware images to find the contained third-party libraries and compared the results. We also compared publicly-known vulnerabilities of those libraries and the device models with the ones SAFER found. Ultimately, we verified the results of the scoring component by using the previously-gathered information as input and validated the predictions made by SAFER.

\subsection{Deployment}
We deployed SAFER in a multinational non-academic research organization that has a large-scale, heterogeneous network infrastructure with more than 300,000 registered network devices. This unique infrastructure enabled us to evaluate SAFER's components in detail. A subset of these 300,000 devices serve general purpose services (e.g. e-mail) or enable file sharing. More specialized embedded devices measure temperature, stream video feeds, enable phone calls, or let users print their documents. The organization has 1,000 registered embedded devices which are distributed in different technical and general-purpose networks. 
SAFER successfully identified 1,000 devices and calculated risk metrics for more than 665 embedded devices.
Moreover, due to vendors reusing the same firmware image for multiple device models, SAFER is able to assess even more device models than currently implemented.
As the organization does not want to impose restrictions, employees are free to register both their business and private devices (BYOD policy). When doing so, users are required to keep their connected devices safe. 
In order to enable employees to adopt safe practices, SAFER is designed to support network users in assessing the security and risks of their devices.

If employees manage multiple devices or even network subnets of the organization and want to assess all those devices at a glance, they are capable of doing so. SAFER implements a feature to assess multiple devices and shows them in a list view with concise information about their security risks.
Access to SAFER within the organization is currently limited to the computer security team. We consider this study as a required step in the process of making it available to all personnel. We recognize that SAFER will have a wide impact on device selections and sanctions within the research organization. Thus, we need to thoroughly understand how professional users interact with such a device risk assessment tool before we roll out SAFER for the entire personnel.

\subsection{SAFER User Interfaces}
We developed two device risk representation views that we refer to as \textit{guided} and \textit{rich} versions (see Figure \ref{fig:figure_dashboard_combined}). Both versions communicate the same key information: the device risk level based on the identified firmware, vulnerability and patch trends for the particular model, as well as any exceptional risks. 
The indicated colors range from green, over yellow, to red. Green refers to none or minor risks, yellow as medium risks, and red indicates high risks.

\begin{figure}
  \centering
  \includegraphics[width=0.7\columnwidth]{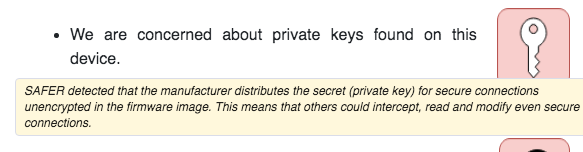}
  \caption{
  Tooltips provide additional information about SAFER's assessments.}~\label{fig:tooltip}
\end{figure}

In the following, we detail both versions of the user interface.

\subsubsection{Version A: Guided}
The \textit{guided} version is depicted on the left of Figure \ref{fig:figure_dashboard_combined}. This version uses a traffic light to reflect the overall risk assessment. In case of the smart kettle, the first sentence indicates that the device poses a high risk for the infrastructure. This information is based on the highest severity \cite{cvssRating} of un-patched vulnerabilities for third-party software which SAFER found in the identified firmware image. SAFER requests all vulnerability information using public vulnerability sources like Mitre \cite{Mitre2019CVE}. The upper red icon indicates that SAFER found several un-patched vulnerabilities in multiple firmware versions of this particular model. 
The second red icon indicates that SAFER found cryptographic key material within the identified firmware. This could allow attackers to intercept secure connections. When users click on or hover over the icons on the right-hand side, SAFER displays explanatory tooltips, as shown in Figure \ref{fig:tooltip}.

\subsubsection{Version B: Rich}
SAFER's \textit{rich} version is depicted on the right of Figure \ref{fig:figure_dashboard_combined}. Two interface components provide an overview of the overall \textit{device risk score} and \textit{future risk estimation}. To learn more about the underlying data and assessment criteria, users can expand those two boxes. Figure \ref{fig:figure_dashboard_combined} shows that the user is reviewing additional information on the \textit{Future Risk Estimation}. For the \textit{device risk score}, we display a table of reports on Common Vulnerabilities and Exposures (CVE) \cite{Mitre2019CVE}, along with the exceptional risks. This table details CVEs, indicating, for example, the severity of the vulnerability and the probability for exploitation. 
The \textit{future risk estimation} provides information on the \textit{firmware vulnerability trend} and the \textit{model patch trend}. The \textit{future risk estimation} is based on these two trends, which relate to the number of vulnerabilities across all firmware images and the number of available patches. Figure \ref{fig:figure_dashboard_combined} shows a high \textit{firmware vulnerability trend}, indicating that the vendor does not have a habit of releasing firmware updates to fix vulnerabilities. 
The \textit{model patch trend} shows a graph with a maximum of two patches per year which are released by the vendor for this model. Since the number of patches is historically low, SAFER indicates a low patch trend. The combination of \textit{firmware vulnerability trend} and \textit{model patch trend} inform the high-risk assessment of the \textit{future risk estimation}. 
Similar to the icon-based tooltips in Version A, SAFER displays additional information when hovering over one of the sections titles (e.g. Model Patch Trend).

\subsection{SAFER's User Interfaces (UI) in the Wider Context}
One of the strengths of SAFER is the framework's focus on device risk communication that is expected to enable users to take informed decisions. SAFER's user interface provides an extensive set of features, including a user's device overview, device assessment through descriptive text, various categories, traffic lights (Guided), detailed trends (Rich), additional descriptions through tooltips, and a comparison of devices across a specific category that is expected to aid device selection. That way, SAFER's user interface differs from related work that focused on single UI components like risk trees / maps \cite{Huang2016}, color-coded rating tables \cite{Loi2017}, and static labels \cite{emami2019exploring}. In this section, we reflect on differences between SAFER's user interfaces and related work.

Our study was inspired by Emami-Naeini et al.'s~\cite{emami2019exploring, emami2020ask} work on security and privacy labels. Still, there are clear differences between SAFER and the labels. The key difference is that SAFER is a \textit{dynamic} web service that updates device risk assessments based on changing security information. This is not possible to the same degree with the \textit{static} security and privacy labels. In addition, SAFER adds detailed descriptions that are expected to aid users in understanding the consequences of using unsafe devices. The system further provides calls to action. While the labels offer an easy to interpret star rating that builds on a similar rationale as SAFER's traffic lights, they do not provide any detailed insight about the assessment. As we expected that clear descriptions would foster user acceptance of recommendations and device sanctions, we focused on adding detailed descriptions to the user interface.

Ekelhart et al.~\cite{ekelhart2009ontology} introduced a tool for AUtomated Risk and Utility Management (AURUM). AURUM implements information security risk management standards to visualize security-related process models and physical models. The UI of AURUM only displays technical abbreviations along with yellow and green bars that represent the implementation progress of technical measures. In contrast, we designed SAFER to display technical information in a clear and structured manner that we expect to be interpretable by all device users, regardless of their technical expertise.

Khadeer et al.~\cite{khadeer2018educating} introduced a pilot website for their security compliance measurement system for consumer IoT. Their security compliance measurement rating from 0-1000 is divided into four categories, represented as A-F. In our development, we hypothesized that additional information and explanations are needed to inform device users effectively about the security and risks of their devices and to enable them to take informed decisions. Thus, in contrast, SAFER provides additional information and explanations together with visual assessment components like the traffic lights. The authors further referred to the Trust Framework\footnote{\url{https://www.internetsociety.org/iot/trust-framework/}} of the Online Trust Alliance\footnote{\url{https://www.internetsociety.org}} which is limited to information about current firmware updates. SAFER goes one step further and makes users aware of how often the vendor released patches and fixes security vulnerabilities.

Alrawi et al.~\cite{alrawi2019sok} contributed a detailed summary for a variety of assessed devices. The tables are a handy tool for security experts who can take them as a technical reference. But, they do not address non-technical users. In particular, the tables do not provide any visual components that indicate overall device assessments. Instead, SAFER represents assessments based on colors and visual components and provides detailed explanations.

\section{Method}

We aimed to build a complete understanding of the impact of device risk assessments on network users in a professional environment. Thus, we focused on recruiting employees with a variety of professional backgrounds. In this section, we provide details about the study participants, as well as the structure of the evaluation sessions. We conclude this section by describing our iterative and collaborative data analysis process.

\begin{table}
    \centering
    \begin{tabular}{c c c c c}
    {\small\textit{Ref}}
    & {\small \textit{Domain}}
    & {\small \textit{
    Experience / Role}}
    & {\small \textit{Gender}}
    & {\small \textit{Group}}\\
    \midrule
    P1 & Scientific Support & 
    PhD Student
    & Female & A (Guided) \\
    P2 & Database Engineer & 
    Junior Staff
    & Male & B (Rich) \\ 
    P3 & Library Support & 
    Thesis Student
    & Female & A (Guided) \\ 
    P4 & Authentication Team & 
    Junior Staff
    & Female & B (Rich) \\ 
    P5 & Science Support Manager & 
    Senior Staff
    & Female & A (Guided) \\
    P6 & Software Developer & 
    Junior Staff
    & Male & B (Rich) \\ 
    P7 & Software Developer & 
    Senior Staff
    & Male & A (Guided) \\ 
    P8 & Software Developer & 
    Senior Staff
    & Female & B (Rich) \\ 
    P9 & Electrical Engineering & 
    PhD Student
    & Male & A (Guided) \\
    P10 & IT Project Management & 
    Team Leader
    & Male & B (Rich) \\ 
    P11 & Physicist & 
    PhD Student
    & Male & A (Guided) \\ 
    P12 & Electrical Engineering & 
    Senior Staff%
    & Female & B (Rich) \\
    P13 & IT Project Management & 
    Senior Staff
    & Female & A (Guided) \\
    P14 & Safety Controls & 
    Team Leader
    & Male & B (Rich) \\
    P15 & HR Professional & 
    Senior Staff
    & Male & A (Guided) \\
    P16 & Controls & 
    Senior Staff
    & Male & A (Guided) \\
    P17 & Science Writer/Outreach & 
    Junior Staff
    & Male & A (Guided) \\ 
    P18 & Controls & 
    Senior Staff
    & Male & B (Rich) \\
    P19 & IT Engineer/Systems admin & 
    Junior Staff
    & Male & A (Guided) \\ 
    P20 & Software Developer & 
    Senior Staff
    & Male & B (Rich) \\ 
    \end{tabular}
    \caption{We recruited employees with a diverse set of professional backgrounds, including technical and non-technical ones. None of the 20 participants worked on computer security related topics.}~\label{tab:table_interviewees}
\end{table}

\subsection{Study Participants}
We recruited study participants with technical and non-technical backgrounds from a wide set of professional domains. 
As depicted in Table \ref{tab:table_interviewees}, participants included engineers, human resource professionals, information scientists, physicists and software developers. 
We further added information about the participants' seniority and roles within the organization. We distinguished between \textit{Thesis Students} (Bachelor / Master), \textit{PhD Students}, \textit{Postdocs}, \textit{Junior} (< 5 years) and \textit{Senior} (>= 5 years) \textit{Staff} members, and \textit{Team Leaders}. 
None of the participants worked within the organization's computer security division. 
As access to SAFER was limited to the computer security team, none of the study participants were able to access the tool before this study. In this context, it is important to stress that we consider this requirements study as a required step in the process of making it available to all personnel.

The recruited participants also represent a highly diverse sample. The 20 participants came from 11 countries. Those are, in alphabetical order: Finland, France, Germany, Greece, India, Italy, Poland, Portugal, Romania, Spain, and the United Kingdom. We assured participants that we would not disclose the nationality of individual employees, as doing so might risk anonymity for several of the study participants, in particular for those coming from small or underrepresented nations or working in small units. We placed high value on guaranteeing anonymity, as we invited employees to talk freely about their practices related to the use of network devices. We considered this culturally and professionally diverse sample a valuable asset in building an empirical understanding of the impact of device risk assessments and the overall attitude towards security considerations related to connected devices in work environments.

The participants' mean age was 32 years (SD = 6.6 years; range: 26 to 56 years old). As depicted in Table \ref{tab:table_interviewees}, seven female and thirteen male employees took part in the study. All sessions were conducted in English. All participants spoke English. We randomly assigned participants to two groups, based on the version of SAFER that they would explore. Participants in group A used the \textit{guided} device risk version, as depicted on the left of Figure \ref{fig:figure_dashboard_combined}. Group B participants experienced the \textit{rich} version, shown on the right of Figure \ref{fig:figure_dashboard_combined}.

\subsection{Study Protocol}

We conducted mixed-method evaluations based on the protocol described in this section. The full protocol is also available as supplementary material.

First, we assessed the participants' general relation to technology and technology use. To do so, we used three subscales proposed by McKnight et al. \cite{mcknight2011trust}: \textit{Faith in General Technology}, \textit{Trusting Stance -- General Technology} and an adapted version of \textit{Situational Normality}. McKnight et al. use the example of spreadsheet products in their Situational Normality subscale. The following is an example of how we modified those items: "I always feel confident that the right things will happen when I use spreadsheet products" (McKnight et al.); "I always feel confident that the right things will happen when I connect devices to the (organization's) network."
    
We re-used and adapted questions related to purchase behaviours, selection criteria and the general understanding of security, from the openly available study protocol from Emami-Naeini et al. \cite{emami2019exploring}. Here, we asked about devices connected to the organization's network. In particular, we asked about selection criteria and comparisons. We then invited participants to define security of connected devices. Next, we asked them to discuss any prior security-related concerns related to connected devices, both in their home and at work.
    
Next, we sent an e-mail to the study participants. The message was sent from the SAFER system, inviting the user to review risks and vulnerabilities of their connected devices. In order to control that every participant reviewed the same device types, we pre-selected devices and asked the participants to imagine those devices to be their own. As shown in Table \ref{tab:devices}, we selected three representative private devices and three representative business devices. We aimed to select a set of devices that all network users could relate to, yet covering a broad range of device types. In order to provide a controlled exposure to a wide range of vulnerabilities and risk assessments, the developers of the SAFER framework defined risk characteristics for those devices based on most common vulnerabilities and data. 
    
Half of the participants were directed to the \textit{guided (Version A)} device pages (see Figure \ref{fig:figure_dashboard_combined} (left)). The other half was directed to the \textit{detailed (Version B)} device pages (see Figure \ref{fig:figure_dashboard_combined} (right)). After each of the six device reviews, we invited the participants to respond to a questionnaire. This questionnaire was based on two more subscales proposed by McKnight et al. \cite{mcknight2011trust}. The \textit{Specific Technology - \textbf{Functionality}} subscale was used to understand how well the SAFER presentation supported the assessment of device security. In order to assess how suitable the device pages are in providing sufficient guidance in the device risk assessment, we also used the \textit{Specific Technology - \textbf{Helpfulness}} subscale. To better understand SAFER's impact on how users evaluate devices and risks, we added the following three statements:

    \begin{itemize}
        \item \textbf{Risk concern}: I am concerned about the risk that this device poses to the computer security of (my organization).
        \item Willingness for \textbf{future monitoring}: I feel that I need to carefully monitor the security evaluation of this device in the future.
        \item \textbf{Removal acceptance}: I am willing to disconnect this device from the (organization's) network in order to reduce the overall risk for the (organization's) computer security.
    \end{itemize}
    
    All questionnaire items were based on a 7-point Likert scale. 
    
    We concluded this part of the study by asking about most important information provided by SAFER. And information that the participants were missing.
    
    Finally, we invited participants to imagine that they would want to acquire one new private device and one new business device. They used the SAFER device category search mechanisms to find smartphones (private) and network-attached storages (NAS) (business). Each query returned three devices of the corresponding category. One green, one yellow and one red device. Participants from Group A stayed with SAFER's \textit{guided} pages. And Group B participants reviewed three \textit{detailed} smartphone assessments and three \textit{detailed} NAS pages. After each of the two category selection scenarios, we invited the participants to rate agreement to the following statements:
    
    \begin{itemize}
        \item \textbf{Consultation}: I would definitely review this device selection page before buying the next (private smartphone / network-attached storage (NAS)). 
        \item \textbf{Decision Influence}: My decision to buy a (private smartphone / network-attached storage (NAS)) would heavily depend on the security risk evaluation of SAFER.
        \item \textbf{Non-listed devices}: I would not buy a (private smartphone / network-attached storage (NAS)) if SAFER did not offer a security risk evaluation of this device.
    \end{itemize}

\begin{table}
    \centering
    \bgroup
    \def\arraystretch{1.2}%
    \begin{tabular}{r|cc}
        & Device Type & Device Risk \\
        \hline
        Private\_1 & E-Book Reader & Low \\
        Private\_2 & Smartphone & Medium \\
        Private\_3 & Smart Kettle & High \\
        Business\_1 & CCTV & Low \\
        Business\_2 & Connected Storage (NAS) & Medium \\
        Business\_3 & Printer & High \\
    \end{tabular}
    \egroup
    \caption{We asked participants to review SAFER's device risk assessments of six devices: three privately owned devices, and three devices provided by the organization.}
    \label{tab:devices}
\end{table}

\subsection{Qualitative Data Analysis}

We collected 11.8 hours of audio-recordings during the evaluation sessions. We transcribed the recordings, used Atlas.ti data analysis software to analyze and code the transcriptions, and performed Thematic Analysis \cite{Blandford:2222613} to identify emerging themes. Initially, two of the authors open coded the first two interview transcriptions. Afterwards, they discussed and merged their codes. The remaining transcriptions were coded by one author based on this resulting code tree. In total, we created 102 codes based on 247 quotations. Finally, the authors collaboratively discussed, adapted and merged code groups, resulting in 13 such groups. Based on those, 
we derived the three themes \textsc{Practices and Concerns}, \textsc{Communication}, and \textsc{Trust}.
For example, the theme \textsc{Trust} is strongly based on the code groups \textit{Mismatch}, \textit{Additional Resources and Mechanisms}, \textit{Richness of information}, \textit{Concerns}, \textit{
Impact / Criticality / Complexity}, and \textit{Trust}. The complete Atlas.ti code group report is available as supplementary material. 

\section{Quantitative Results}
In this section, we present results from the quantitative data analysis of our questionnaires 
using R 3.6.3~\cite{rcore}, lme4 1.1-21~\cite{lme4}\footnote{Additional libraries: optimx 2020-4.2~\cite{optimx}} and multcomp 1.4-13~\cite{multcomp}. 
All questionnaire responses are available as supplementary material.

We performed a linear mixed effects analysis to evaluate the impact of the device risk presentation on the reported scores from our participants: the two \emph{Specific Technology} subscales "Functionality" and "Helpfulness" as well as our added statements on "risk concern", "future monitoring", and "removal acceptance". 
All questionnaire items are based on a 7-point Likert scale. Given the limited number of samples, we note that the interpretability of these models is restricted. Hence, this analysis serves to complement our qualitative analysis. Respective pointers are included. 

\begin{figure}
    \centering
    \includegraphics[width=.45\columnwidth]{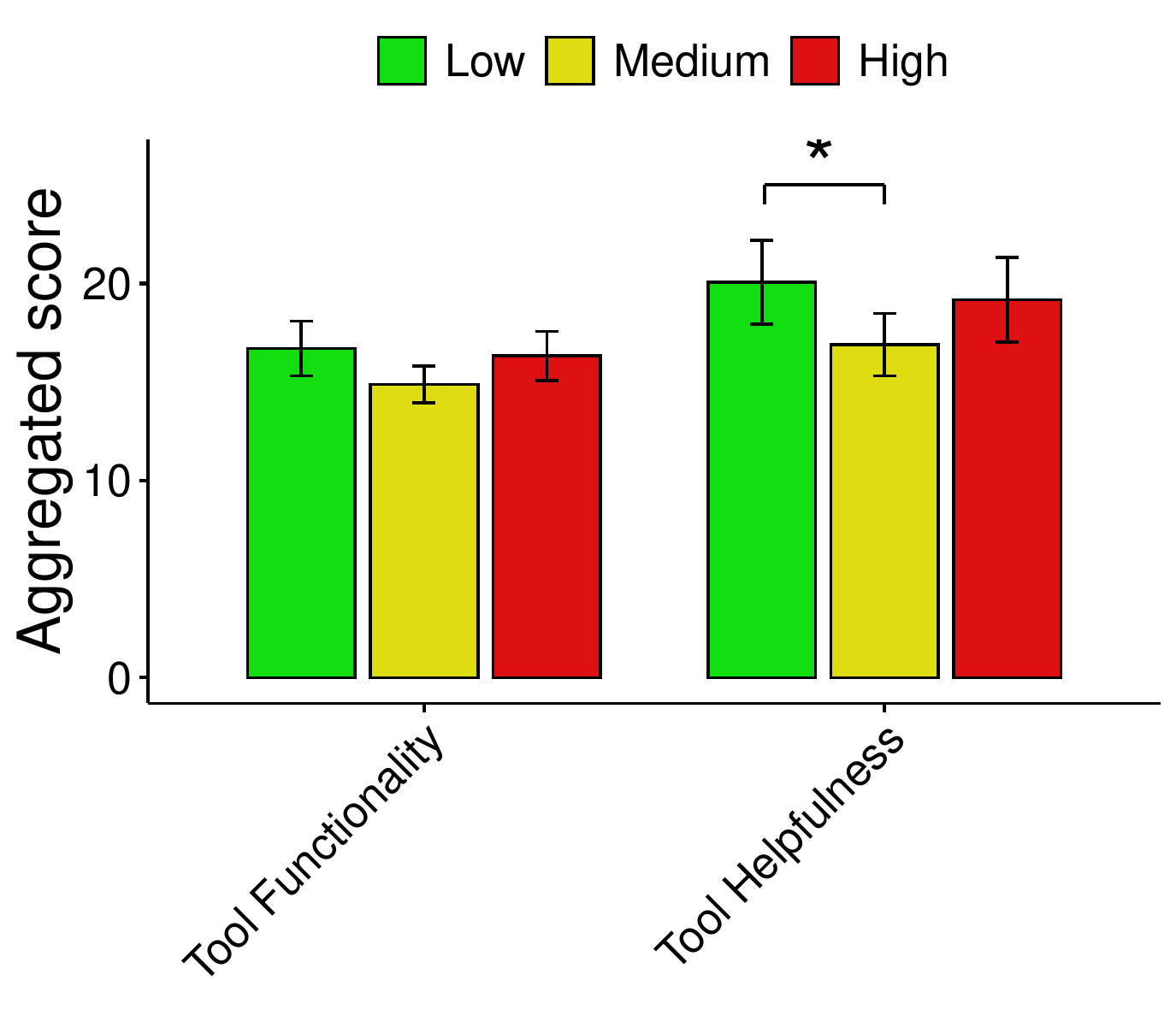}~
    \includegraphics[width=.6\columnwidth]{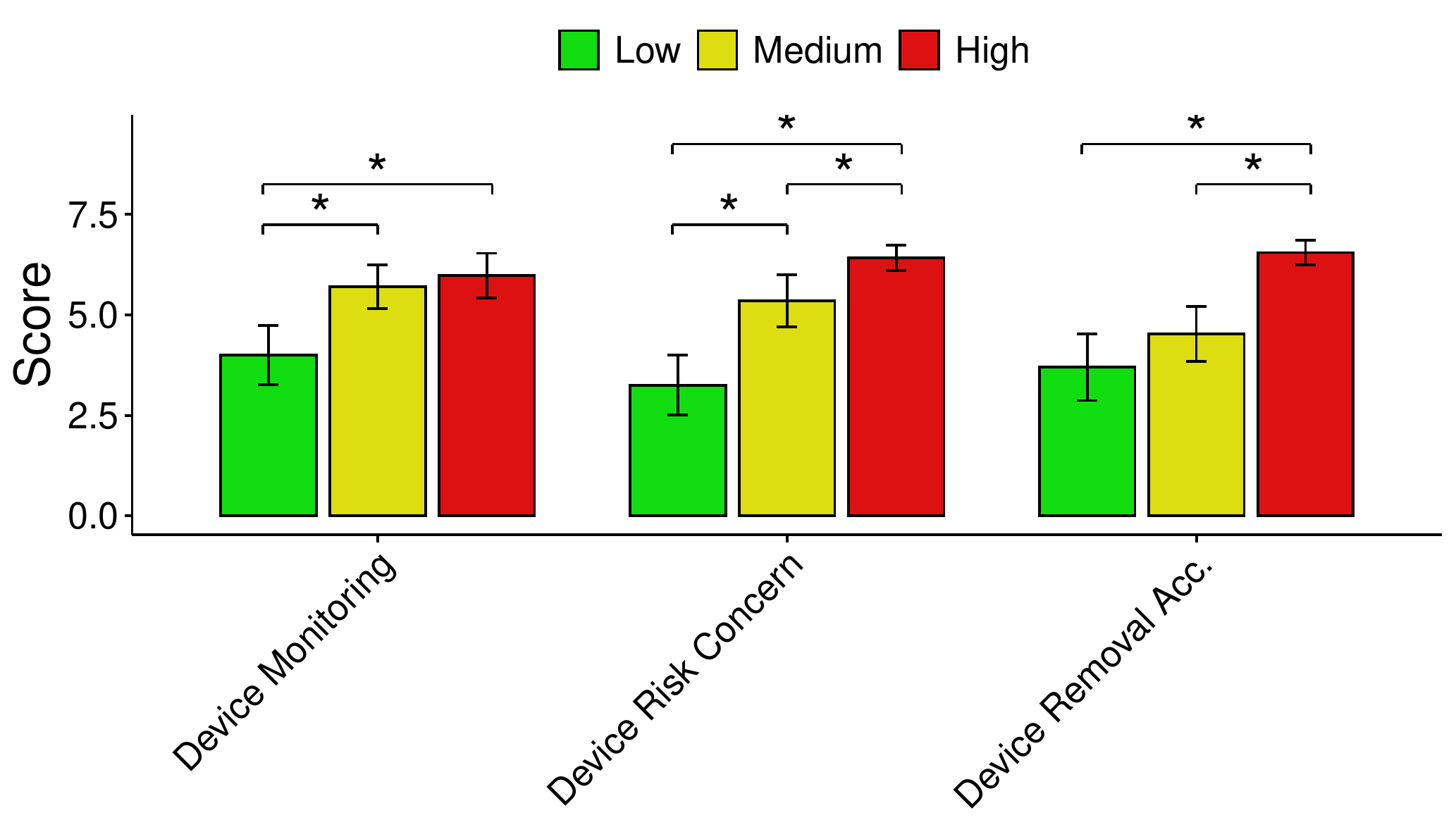}
    \caption{Scores grouped by "Device risk assessment". Error bars show mean confidence intervals. Bonferroni corrected significant differences ($p < .017$) between conditions are marked with *.}
    \label{fig:risk}
\end{figure}

\begin{figure}
    \includegraphics[width=.45\columnwidth]{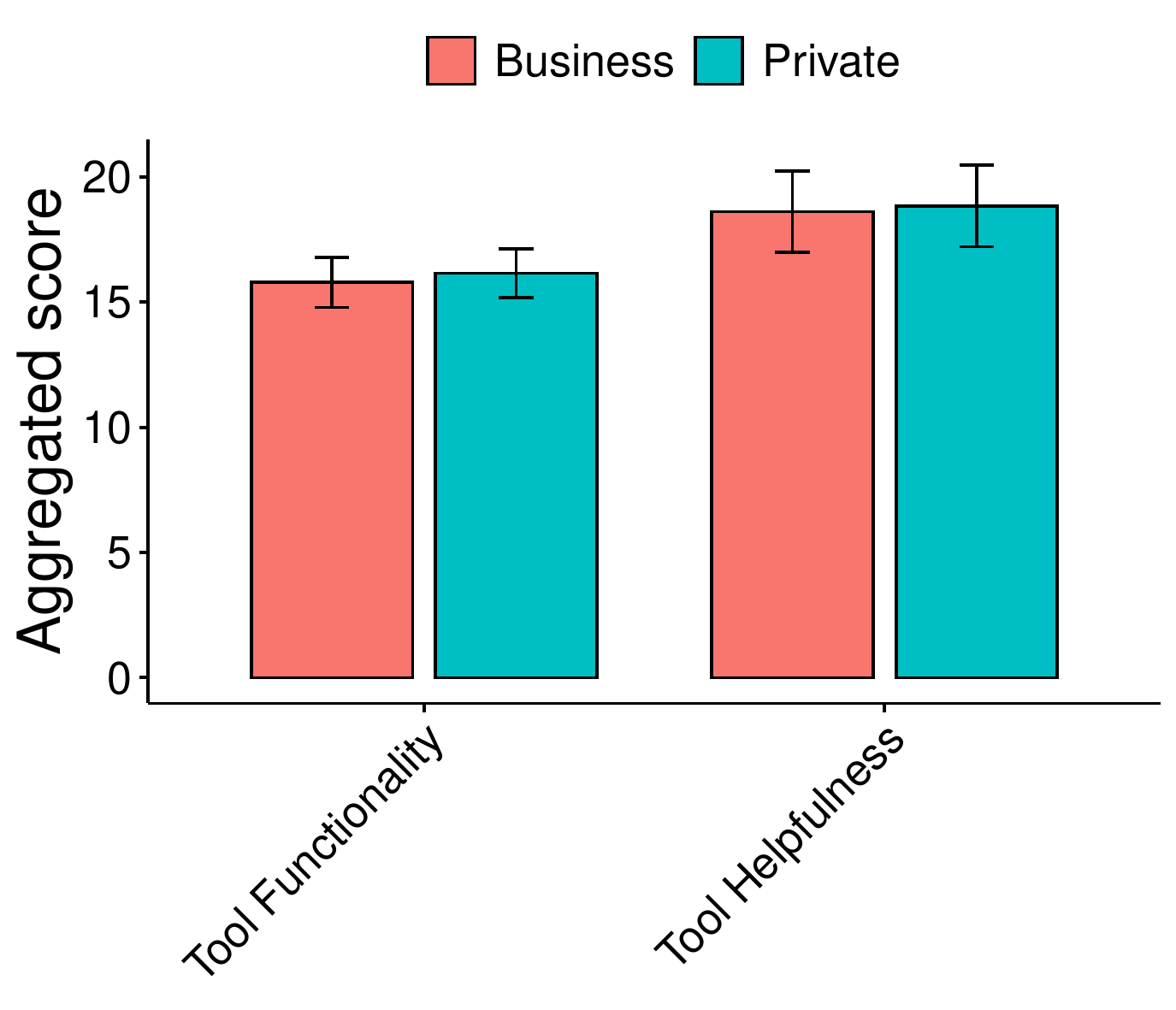}~
    \includegraphics[width=.6\columnwidth]{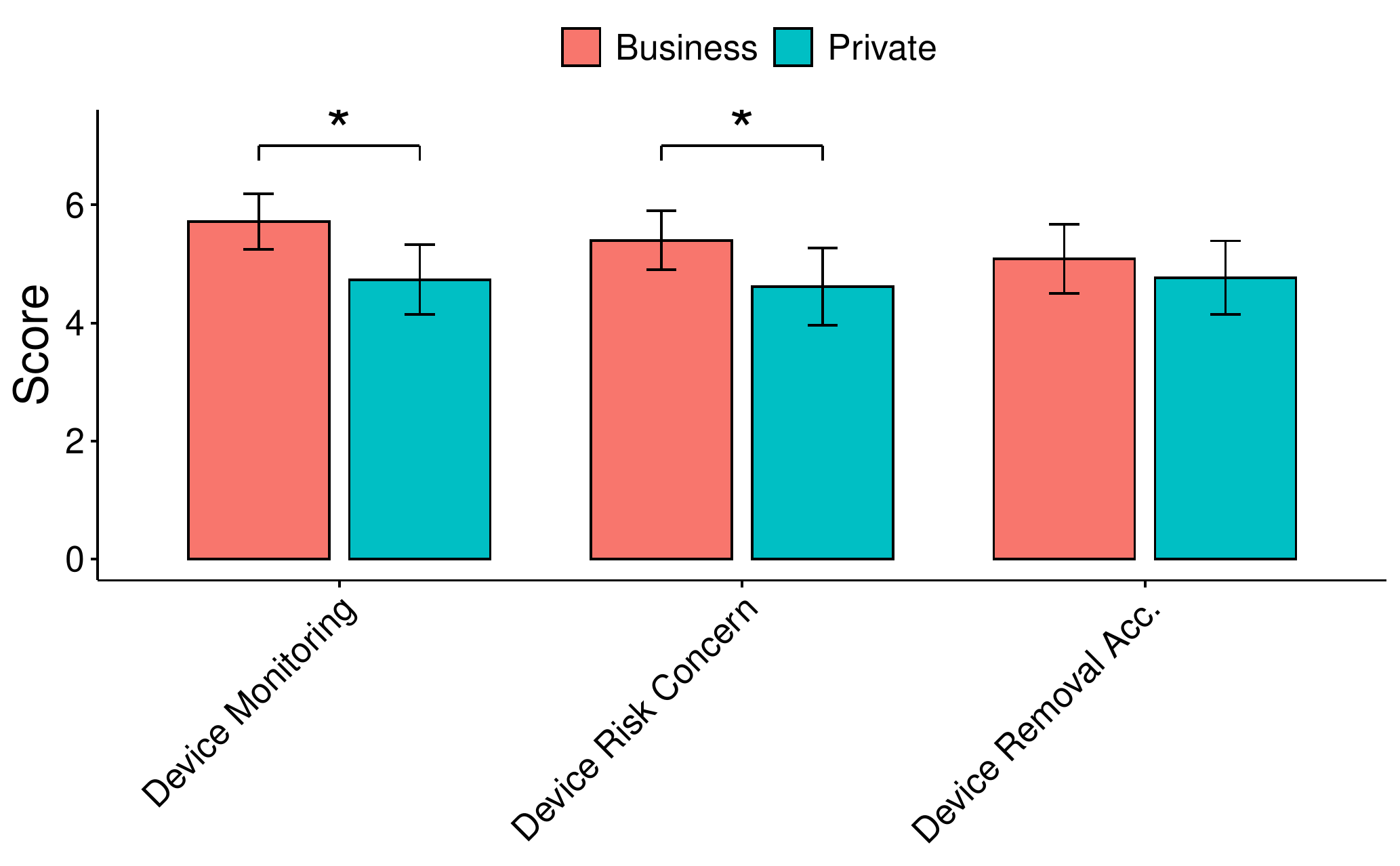}
    \caption{Scores grouped by "Device category". Error bars show mean confidence intervals. Significant differences ($p<.05$) between conditions are marked with *.}
    \label{fig:category}
\end{figure}

Apart from having access to different representations of device risks, namely "Guided" and "Rich", we identified the general availability of having a tool to assess risk importance. Hence, we ran an analysis for the main effect of "Device risk representation" (levels "Guided" and "Rich"), as well as for "Device risk assessment" (
Fig. \ref{fig:risk}) and "Device category" (
Fig. \ref{fig:category}) in separate models.

For this, we created five models for every main effect: one for each of the reported scores. For the first series of models, we entered "Device risk representation" as fixed effect apart the participants' reported "Faith in general technology", "Trusting stance", "Situational normality", "Device risk assessment", and "Device Category" (no interaction effects). As random effects, we had intercepts for participants, as well as by-participants random slopes for the effect of "Device risk representation", "Device Risk assessment", and "Device Category", respectively. 
To compare models, we compare the full model with the main effect in question against the model without the main effect in question (null model). We compare these models based on their AIC\footnote{Akaike information criterion} and use likelihood ratio tests to assess if the added effect significantly improves model fit ($p < .05$). 
For "Device Risk representation", no significant effects were found. 
This means that the presentation (Guided vs. Rich) did not significantly impact perceive tool functionality or helpfulness. It did also not have a significant impact on users' perceptions of device risks. Our analysis of "Device risk assessment" and "Device category", as well as our extensive qualitative findings further explain this finding. 

We repeated 
the 
above analysis with both "Device risk assessment" and "Device category" in an analog fashion. 
For "Device risk assessment" all scores were significantly affected, cf. Table \ref{tab:device_risk_stats}. For "Device category", only "Device monitoring" (${\chi}^2(1) = 7.12, p < 0.05$) and "Device risk concern" (${\chi}^2(1) = 7.22, p < 0.05$) were significantly affected as shown in Fig. \ref{fig:category}. 
Significant results of pair-wise comparisons (adjusted using Bonferroni correction\footnote{Significance level of $p <.017$}) of means for the factor "Device risk assessment" using Tukey contrasts are illustrated in Fig. \ref{fig:risk}. 
In contrast to the risk representation, those results show that both SAFER's risk assessment (Low / Medium / High) and the device category (Business / Private) significantly affect users' interpretations of the tools and devices. In the context of our study, this indicates that the impact of SAFER in the organization is less dependent on selecting the best risk representation. Instead, it is important to provide users with a suitable overview of device risks, possibly in the categories Low / Medium / High, and an indication whether listed devices are business or private devices. 

\begin{table}
    \centering
    \begin{tabular}{lll}
        \toprule
        Tool functionality & ${\chi}^2(2) = 6.51$ & $p < 0.05$\\
        Tool helpfulness & ${\chi}^2(2) = 8.41$ & $p < 0.05$\\
        Device monitoring & ${\chi}^2(2) = 18.03$ & $p < 0.05$\\
        Device risk concern &${\chi}^2(2) = 27.68$ & $p < 0.05$\\
        Device removal accept. & ${\chi}^2(2) = 24.56$ & $p < 0.05$\\
        \bottomrule
    \end{tabular}
    \caption{Results for quantitative analysis for "Device risk assessment" as main effect.}
    \label{tab:device_risk_stats}
\end{table}

Fig. \ref{fig:risk} depicts participants' assessment of SAFER's functionality and helpfulness, as well as users' willingness to monitor and remove devices, and their risk concern for devices. Notably, users rated SAFER's functionality similar for all devices and across all device risk levels. We noted a slightly stronger difference in perceived helpfulness. Here, participants voted helpfulness significantly higher for devices that have low risk assessment compared to devices with medium risk assessment (cf. Fig. \ref{fig:risk}). 
As our qualitative findings show, this is likely due to the fact that users were occasionally undecided about actions that needed to be taken in case of Medium risk devices. Rather, they asked for more binary assessments (Green / Red or Low / High). 

Concerning the devices, users perceived a strong need to monitor the medium and high risk devices (both significantly higher than for low devices), as depicted by Fig.~\ref{fig:risk}. For monitoring, the device category also influenced the participants' rating, being significantly higher for business devices (cf. Fig.~\ref{fig:category}). For example, participants found it much more important to monitor the CCTV camera ($5.1$), than the e-book reader ($2.9$). This is notable, as the risk assessment and the calls-for-action are identical for those two low-risk devices. 
We explain this effect through our qualitative findings. Here, we noted that for some devices, users' personal perception of the devices mismatched with SAFER's assessment. This is particularly true for the CCTV camera that, although assessed as a low-risk device, was judged more critical because of personal aversion against those devices. This is a valuable finding for the design of systems that communicate device risks and assess device security. We discuss implications in detail in the Discussion section. 

For the participants' ratings on risk concern, the differences between the assessed risk levels are even higher. Devices assessed with high risk are a major concern, followed by medium and low risk devices. All differences are significant. Again, business devices are of higher concern for the participants.

\begin{figure}
    \includegraphics[width=0.75\columnwidth]{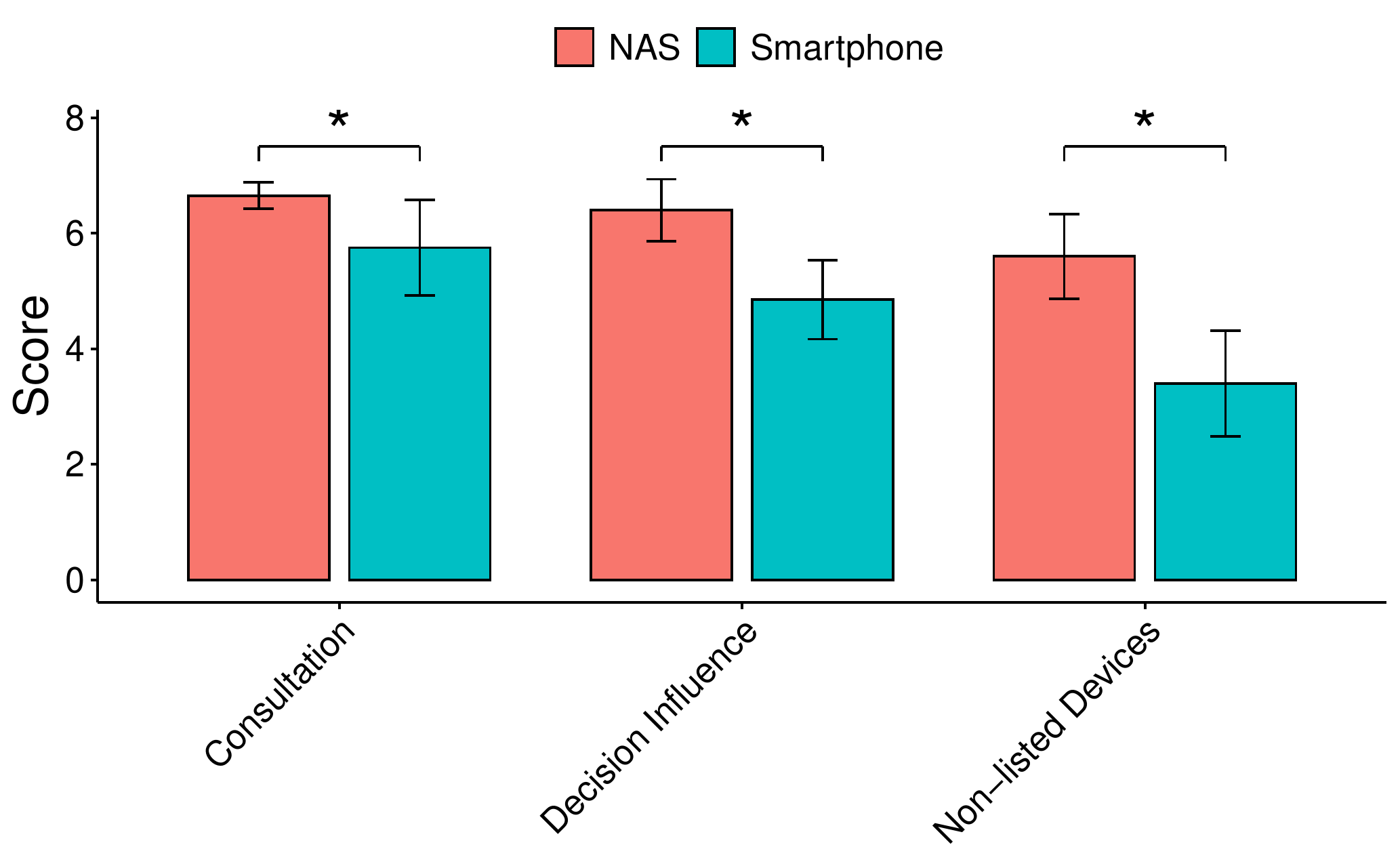}
    \caption{Participants want to consult SAFER when they buy a new device. However, SAFER has a much stronger impact on the selection of the business device (NAS), compared to the private one (smartphone). 
    Significant differences between conditions are marked with *.}
    \label{fig:selection_impact}
\end{figure}

At the end of the evaluation sessions, we asked participants to imagine that they would want to buy one new private device (smartphone) and one new business device (network-attached storage NAS). We asked them to use SAFER to review device risk assessments of devices of those two categories. Here, we find a strong difference on SAFER's impact on device selection. As shown in Figure \ref{fig:selection_impact}, participants --- for both device types --- clearly stated that they would consult the tool before buying a device (Smartphone: $\bar{x}=5.75$; NAS: $\bar{x}=6.65$). However, the decision influence ($\bar{x}=4.85$) and the willingness to buy only devices that were assessed by SAFER decreased rapidly for the private smartphone ($\bar{x}=3.4$). For the business device, they stayed both at a high level (decision influence: $\bar{x}=6.4$ and non-listed devices: $\bar{x}=5.6$).
A Wilcoxon signed ranked test confirmed a significant difference of the device (business NAS, private Smartphone) with regard to whether users would consult SAFER ($V=42, p < .05, r=.54$), whether it influenced their decision ($V=120, p<.05, r=.80$) and their willingness to buy devices not listed within SAFER ($V=120, p<.05, r=.80$). As our qualitative findings show, device security is only one of several criteria in the selection of private devices. Instead, risk assessment becomes a key criterion in the selection of business devices when such assessments are available. This is a valuable finding for organizations and enterprises planning on introducing systems that communicate device security as part of a selection process. 

\section{Qualitative Results}

Based on our qualitative data analysis, we present three themes that describe practices, requirements and needs related to the design of tools that provide device risk assessments: \textsc{Practices and Concerns}, \textsc{Communication}, and \textsc{Trust}. The themes provide valuable insight into users' interaction with SAFER and allow to reason about the results.

\subsection{Practices and Concerns}

Almost all participants reported being concerned about the security of their devices. Most referred to concerns about privacy violations and loss of confidential data. As active security measures, several participants reported using anti-virus software. In general, participants reported using between three and six devices in the organization's network. Some of the participants registered a far greater number of devices that were used by multiple users. For example, P10 is responsible for 50 devices. P14 for several hundred devices:

\begin{quote}
"The control team or external contractor teams use them." (P14)

"I am registered as the responsible person for the hardware of some devices. So for instance for the monitoring infrastructure, all administrators from the group have access to these machines." (P10)
\end{quote}

Informants also told us about how they got their business devices. Few of the participants said that they could chose at least one of the devices that were given to them. Most mentioned that their supervisors either ordered a device before their first working day or that they inherited devices: \textit{"I got the workstation from my team. Some of them were spare ones."} (P17)

Besides thinking about the impact on their own privacy and the vulnerability of their data, some participants even thought about potential consequences on the overall network:

\begin{quote}
"A device could be a loophole to connect to the (organization's) network. But I do not evaluate the security of the devices I own / connect." (P12)
\end{quote}

This reflects a common perception of many participants, who highlighted repeatedly that they were not able to assess the security of their devices. Exemplarily, (P11) said: \textit{"I have a strange feeling about my phone. However, I still connected my phone to the network, because I am also not sure if a different phone is better or not in sense of security."}

\subsection{Communication}

Our data analysis showed that it is essential to communicate risks in a clear way. Most participants found provided information and especially the color-coded assessments useful:

\begin{quote}
"The color and score concerns me a lot. A high risk would mean that the device can be dangerous for the network, the data and myself." (P4)

"The overview of the traffic lights is enough, I do not need to evaluate further." (P16)
\end{quote}

Those statements illustrate how communicating device risk categories directly impacts risk concern. Although the risk category and the traffic light color proved to be the most prominent and important information in device evaluation, we found that wording needs to be carefully considered in many cases. For example, P13 stressed that the system might exaggerate in case of the red, high-risk kettle, as the descriptive text does not convey a stronger concern:

\begin{quote}
"And since it says 'we are concerned' - so, it's not so strong, it's not like 'don't use it!'... so, I would say that maybe this is - this high risk is a mistake. And it's an exaggeration of the system because it's not updated frequently for example."
\end{quote}

This statement illustrates the user's interpretation of the wording, as well as a mismatch between SAFER's assessment and the user's understanding of risks. In general, participants asked for more guidance, in particular related to concrete actions that they should take. They imagined both human and technological support:

\begin{quote}
"I think in general it would be nice to have some guidance. A back office that I can contact. [...] I would want to have more information about the problems that lie behind the private keys e.g. Behind the public vulnerability report." (P15)

"Most of the devices you have showed mentioned the private keys, the potential risk et cetera. Maybe it should tell me how to minimize the risk. The yellow device indicated firmware updates, ok, so I update them regularly. But I don't know what to do if private keys were found and how bad it is." (P11)
\end{quote}

We noted that the device risk category and the color scheme represent key indicators in the device assessment. However, in some cases they irritated users who consequently asked for more \textit{binary} actions and ratings that do not require further evaluation. This likely contributed to participants rating overall \textit{helpfulness} lower than \textit{functionality}:

\begin{quote}
"I felt like, also from my reaction, when it's yellow, I don't know exactly what to do. And it's a bit of a grey zone. I would prefer it to be either green. You can use it. Or red, you should not use it. If it's yellow, I would prefer it to be red, actually. Because, now I am a bit confused." (P13)
\end{quote}

Overall, participants found SAFER functional and important to use. They even asked for the system to be integrated into the wider service architecture of the organization. Several users expressed their desire to keep up-dated and notified by the system. This would enable a reliable and simple exchange of crucial information: 

\begin{quote}
"I am missing an option to get informed about the device, after I scanned the device the first time. So if new vulnerabilities are known, I will get an email and be aware of it and don't need to scan my device every month." (P2)
"Would be nice if SAFER shows, if a new firmware than the current one is available. Additionally, it would be nice if one could subscribe to a reminder if a new firmware comes out." (P7)
"I would like to have a way to open a ticket to the (organization's) security team in order to coordinate upcoming steps based on the risk of the device." (P1)
\end{quote}

\subsection{Trust}

Users have their own perceptions of the impact and criticality of devices. This perception and the perceived uses of a particular device influence users' acceptance of sanctions and their willingness to monitor devices. The following are representative statements that relate to deeply routed concerns for particular types of devices, professional attitudes, and values of connected devices in organizational networks. 

\begin{quote}
"Some things that changed, since it is a camera. There is no risk right now, but it is a camera and a device I would carefully monitor, since it can be a risk for privacy." (P4)

"The decision of a professional and private device would differ. I should think of the workplace values and not necessarily my values which refers to security and privacy. My own (lax) view of security and privacy should not be applied, when using the device in the workplace." (P9)

"It also depends on the device. This one is a kettle, so it sounds bad, but it is a kettle and unimportant to be on the network. It would also absolutely depend how important the device is for me and if I absolutely need it for my work in order to make the point to disconnect it or not." (P8)
\end{quote}

Notably, users' concerns and general attitude towards certain types of devices influenced and even contradicted SAFER's risk assessment, as illustrated by P4's statement. We find that this mismatch is based on privacy concerns (e.g. CCTV cameras) in most cases. In addition, when the perceived device complexity and impact do not match the device risk assessment of SAFER, mismatch impacts trust in the entire system. For example, P13 made such considerations when first reviewing the red kettle and then the red printer:

\begin{quote}
"So, I see high risk. And I am thinking what could possibly be the high risk? [...] I don't know how much I can trust this system, basically. So, I feel that I would ask around and say 'what do you think about this - is this a system I can trust, or not?' I am a bit conflicted." (Kettle, red) \newline

"It's so funny, I can't believe that my impression is different. That's crazy. Now, for example, I feel that the tool, of course, can assist to assess the security of the device. Before I thought that the system cannot judge." (Printer, red)
\end{quote}

\section{Discussion}

In this section, we describe how device risk assessment tools enable network users to make informed decisions about their connected devices. We discuss why such support is particularly relevant for professional users, how to design for effective support, and how the professional environment fosters adoption.

\subsection{Work Environment}

Private consumers usually select their devices themselves. To do so, they consider a wide set of criteria, including features and price. Instead, device security and privacy criteria are often not instrumental in the selection of devices as users often do not have knowledge or information about their security \cite{emami2019exploring}. Our findings in a professional context show that SAFER's risk assessments provoked significantly higher concern for business devices, as well as increased willingness to monitor those professional devices if they are considered vulnerable. However, our findings related to the \textsc{Practices and Concerns} theme show that professional users often do not have the freedom to select business devices. Their \textbf{device selection might be restricted} by corporate catalogues. In addition, \textbf{professional devices are often be pre-selected} by their supervisors \textbf{or simply inherited}. This is an aspect of our study that differs strongly from the influential work of Emami-Naeini et al. \cite{emami2019exploring}. In addition, we find that: some employees are \textbf{responsible for devices that they do not use themselves}; devices in work environments not only manage private information, but often contain \textbf{critical and confidential} information that can jeopardize an organization's internal and external reputation and success; and \textbf{employees worry about sanctions} related to unsafe practices and data loss. In fact, using SAFER's support in the selection of one new private smartphone and one new business NAS, participants showed more concern in the selection of the corporate device. Our findings showed that the study participants considered SAFER a valuable tool in reviewing the security and risk assessment of connected devices.

\subsection{Usability}

Kirlappos and Sasse \cite{kirlappos2014usable} emphasized that an important aspect of compliance is "trusting employees to 'do what's right' for security." However, our findings suggest that users often do not have the tools to assess \textit{what's right}. As such, we consider the development of SAFER an important step towards establishing \textit{usability} in computer security, which the authors defined as "improving employee ability to behave in a trustworthy way." 

Our findings in the \textsc{Communication} theme show that we have to consider various requirements in the design of device risk assessment tools, with particular regard to the risk communication. Most participants pointed out that the color-based device risk assessment conveyed most important information. This is also reflected in the quantitative results that show a significantly higher risk concern and willingness to monitor medium and high risk devices as compared to low risk devices. This also explains why we did not find significant differences in tool helpfulness and functionality between the two versions of SAFER, as both provide a color-coded assessment that is directly linked to the low, medium, and high risk assessment. Our hypothesis that the guided version would be perceived more helpful because of the descriptive text blocks was not confirmed. Based on our findings, this is likely due to the fact that users perceived a mismatch between some of the wordings (e.g. a rather weak 'we are concerned') and the overall assessment of the device (e.g. high risk). Thus, \textbf{designers of risk assessment and communication tools need to carefully consider the implications of wording in relation to the overall assessment.}

Our findings show that \textbf{the design of device risk assessment technology needs to consider the delicate relationship between risk communication and user perceptions of device complexity.} Users' risk concern and removal acceptance for the low-risk CCTV camera demonstrates this very clearly. As could be expected, users showed little concern for the low-risk E-Book reader. But, more concern for the low-risk CCTV camera. Here, the low-risk assessment on SAFER does not match users' perceptions of the overall risk on privacy for such a device. In the findings section, this is nicely reflected by P4's representative statement regarding CCTV cameras. Similarly, participants found it difficult to imagine how a simple device like a smart kettle could be assessed as high-risk. This is due to the fact that users did not consider the smart kettle as a complex device, and did not see the potential harm, as the device does not store confidential data. Thus, \textbf{we need to carefully explain risk assessments with a particular focus on the difference between the risk assessment of the firmware of a device and the implications for privacy}. This is crucial, as our participants remarked that \textbf{a mismatch between SAFER's assessment and users' perceptions of device complexity impact their overall \textsc{Trust} in the tool.}

\subsection{Adoption}

Emami-Naeni et al. \cite{emami2019exploring} discussed how security and privacy labels on end consumer products can support security awareness and security assessment among private consumers. SAFER enables similar mechanisms, while providing a more dynamic assessment of IoT device risks. SAFER not only informs the selection and purchase of devices. Instead, it provides risk assessments to device users and owners during the entire time a device is connected to the network. When device risk assessments change, SAFER could even notify corresponding users within the organization. This requires a clear strategy for interfacing with other organizational systems. In order to foster the adoption of tools like SAFER in corporate environments, \textbf{designers and computer security teams should consider the integration of device risk assessment tools into already existing organizational service architectures}. Our study participants, for example, asked for integration of SAFER into the organization's ticketing system. This request reflects a wider call for additional guidance. In particular, study participants stressed that they did not know what to do for medium risk devices. This is also reflected in our quantitative results: Participants rated tool helpfulness significantly higher for low risk devices as compared to medium risk devices. Several participants expressed that a binary assessment of device risks could in fact be more helpful. The call for further guidance relates both to calls-to-action on the SAFER system and for further human support. We must acknowledge that \textbf{guidance will be crucial in the wide-spread adoption of tools like SAFER in corporate environments and developments towards a culture of device security awareness in professional BYOD environments.}

\section{Limitations and Future Work}

We aim to foster the replicability of our work and to provide a base for future research. Thus, we provide several of the study's resources as supplementary material. Those include the evaluation protocol, the Atlas.ti code group report, the questionnaire, and the questionnaire responses. We expect that this will enable future research on the design and evaluation of device risk assessment tools. Both, for private and corporate environments. 

We also want to reflect on the limitations of our study. 
Most importantly, the selection of specific user interface components affected the results of the study. We used a traffic light in SAFER's guided version, to draw on users' familiarity with this everyday metaphor. Future work should systematically investigate and compare additional status components like star ratings or numeric assessment criteria. Such an investigation should also further assess whether users look deeper into the meaning behind the status components or focus solely on visual cues like colors. 
We based the device risk assessments of the six devices on real observations of SAFER and characteristics that the tool is able to assess. However, in order to design a study with a rich set of devices, risks, and comparable device criteria, the device assessments in our study were based on the manual construction of six cases by a computer security engineer. We chose to apply the identical device risk metrics and descriptions to devices of the same risk category. While this is a limitation of the study, we consider it an important approach that supports comparison between private and business devices. In fact, we argue, that this selective approach enabled us to describe implications for the communication of device risk assessments. In particular, regarding mismatches between SAFER's assessment and user's perceptions of device complexity and impact. 

CERN is organized into ten departments and hundreds of groups, sections, and teams. We note as a limitation that recruiting participants from within all departments, groups, sections, and teams was not feasible. We further note that although we did not recruit any computer security experts, participants might have had prior exposure to security-related topics. This might be especially true for informants with a technical background. Still, we want to emphasize that our recruitment strategy has focused on sampling for diversity, both in terms of technical background and departmental structure.

SAFER scans devices under test non-intrusively and prevents aggressive pen-testing which causes errors and fault states. SAFER is designed to fetch device characteristics with as little device interaction as possible. It is an important design decision that SAFER is not designed to be a replacement for security teams. Instead, we consider it a tool that helps organizations' security experts keep track of connected devices, and device users to make informed decisions about their devices. 

We envision that the future development of SAFER will place particular emphasis on reflecting our findings and design implications. We expect that a wider deployment within the research organization will follow this initial user study and corresponding improvements. 
Future work might further explore how device risk assessment tools impact risk awareness and security practices across a wide set of professional networks and relate findings on a large set of assessed devices in a production environment to our initial findings on design implications for risk assessment tools. 
In addition, our understanding of the adoption of such tools in corporate environments would profit from participation of different types of technology users, including large infrastructure managers, employees responsible for corporate device catalogues, and computer security teams.

\section{Conclusion}

This paper presented a systematic study of IoT device security practices, concerns, and design requirements for risk assessment tools in a large multinational research organization. We detailed the development of SAFER, a device risk assessment framework that we evaluated with 20 professionals in a mixed-method study. We presented our findings based on three themes that describe practices, requirements, and needs related to the design of tools that provide device risk assessments: \textsc{Practices and Concerns}, \textsc{Communication}, and \textsc{Trust}. We find that corporate employees are concerned with the security and privacy of the devices that they connect to the organization's network. In particular, because they are often not able to choose their business devices themselves. Those decisions are usually made by their supervisors, or they simply inherit connected devices. However, most of the participants pointed out that they are not able to assess the security of their devices. Our findings show that the participants value the SAFER system as a tool that supports them in making informed decisions about their connected devices. We discussed design implications that relate to the communication of device risks and calls-to-action. In particular, we described the need to better explain the meanings and impacts of device risks, as the study participants reported mismatches between SAFER's device risk assessment and their own perception of the complexity of a device. Finally, we discussed how the integration of device risk assessment tools into the corporate service architecture can ease the adoption of sought device security practices.

\section{Acknowledgments}

This work has been sponsored by the Wolfgang Gentner Programme of the German Federal Ministry of Education and Research.

\bibliographystyle{ACM-Reference-Format}
\bibliography{sample-base}

\end{document}